\documentclass[conference]{IEEEtran}
\IEEEoverridecommandlockouts

\usepackage[cmex10]{amsmath}

\usepackage{enumitem}
\usepackage{graphicx}
\usepackage{color}
\usepackage{amsmath}
\usepackage{mathtools}
\usepackage{multirow}
\usepackage[english]{babel}
\usepackage{blindtext}
\usepackage{algorithm}
\usepackage{algorithmic}
\usepackage{balance}
\usepackage{amsfonts}
\usepackage{bm}
\usepackage{stfloats}
\usepackage{subfig}
\usepackage{amsthm}
\usepackage{amssymb}
\usepackage{setspace}
\usepackage[nosort]{cite}
\usepackage{CJK}
\usepackage{cite}
\usepackage{caption}
\usepackage{comment}
\usepackage{array,multirow}
\usepackage{graphicx}
\newtheorem{lemma}{Lemma}
\newtheorem{theorem}{Theorem}

\theoremstyle{plain}

\usepackage[table]{xcolor}

\def\BibTeX{{\rm B\kern-.05em{\sc i\kern-.025em b}\kern-.08em
    T\kern-.1667em\lower.7ex\hbox{E}\kern-.125emX}}
\begin{document}

\title{An Analytical Model for Coordinated Multi-Satellite Joint Transmission System\\
}

\author{
\IEEEauthorblockN{Xiangyu Li\IEEEauthorrefmark{1,2}, Bodong Shang\IEEEauthorrefmark{1}*}
    \IEEEauthorblockA{
    \IEEEauthorrefmark{1}Eastern Institute for Advanced Study, Eastern Institute of Technology, Ningbo 315200, China\\
    \IEEEauthorrefmark{2}Department of Electronic Engineering, Shanghai Jiao Tong University, Shanghai 200240, China\\
    e-mail: \{xyli, bdshang\}@eitech.edu.cn
    }
}

\maketitle

\begin{abstract}
Satellite communication is one of the key technologies that is enabling next-generation networks. However, the nearest-satellite-supported downlink transmission may not meet a user terminal (UT)'s requirements due to limited signal strength, especially in emergent scenarios. In this paper, we investigate a coordinated multi-satellite (CoMS) joint transmission system from a system-level perspective, where a UT can be served by multiple satellites to improve its quality-of-service (QoS). Furthermore, we analyze the coverage and rate of a typical UT in this joint transmission system. Simulations and numerical results show that the introduced system achieves a higher coverage probability than that of the traditional nearest-satellite-supported network. 
Moreover, a UT's ergodic rate can be maximized by selecting an appropriate elevation angle.
\end{abstract}


\section{Introduction}
Satellite communication (SatCom) is playing an increasingly significant role in today's fifth generation (5G) and beyond 5G (B5G) wireless communication systems. Due to its deployment locations and characteristics in the space environment, the satellite system is able to provide pervasive and seamless network coverage as a complement for terrestrial networks, where unavoidable blockages and limitations can be largely overcome.
Apart from the geostationary orbit (GEO) satellites, low-earth orbit (LEO) satellites, which are deployed below an altitude of $2,000$ km and faced with fewer challenges such as long transmission delay and power consumption \cite{9520380}, can serve as ideal candidates for SatCom.

However, in a nearest-satellite-supported area where higher data rates are required for user terminals (UTs), the current satellite network structure may not be adequate.
The terrestrial UTs may experience poor communication conditions, especially in emergent scenarios or when satellites are in a sparse area, which leads to severe scattering of transmitted signals.
Under such cases, if more satellites join in the transmissions of data, the desired signals will be increased and interference signals will be reduced, while the bandwidth allocated to each UT can be decreased.
Hence, it is imperative to model and analyze a coordinated multi-satellite (CoMS) joint transmission system, where a UT can be simultaneously served by multiple satellites to overcome the limitations of poor desired signal strength and severe co-channel interference in the traditional nearest-satellite-supported network.

Many previous papers have studied SatCom systems. The downlink coverage and rate analysis in the LEO satellite network are investigated in \cite{okati2020downlink}. In \cite{9218989}, satellite gateways are deployed on the ground to serve as a relay between the LEO satellite and UTs. 
The use of Poisson point processes (PPP) is analyzed in \cite{park2022tractable}, where tractable expressions for coverage in downlink satellite networks are obtained. In \cite{10107732}, the authors studies the satellite-to-airplane system where each airplane is served by the nearest satellite.
Related works \cite{liu2020cell,li2021satellite} only use one LEO satellite as a single component out of the terrestrial networks.
Nevertheless, these above works concentrate on a typical UT served only by the nearest LEO satellite instead of the satellite coordination. 
The coverage performance in cooperative LEO satellite networks is investigated in \cite{shang2023coverage}; however, the use of order statistics makes the derivations complex for multiple satellites, and the corresponding achievable data rate is not considered.
The massive multiple-input multiple-output (MIMO)-enabled LEO satellites have been studied in \cite{abdelsadek2023broadband,abdelsadek2022distributed}; however, the location randomness of satellites and UTs in these works is ignored and the network nodes are confined to a certain small area-of-interest.
Moreover, although the non-terrestrial networks (NTNs) are integrated with the cellular communication system for higher coverage and data rate as in \cite{10530195}, where one of the satellites is scheduled to serve UTs, the joint transmission performance itself for multiple LEO satellites remains unknown.

Motivated by the above observations, we derive the analytical expressions and analyze the performance of the CoMS joint transmission system\footnote{The CoMS transmission system is feasible by employing a central base station (BS) to coordinate the simultaneous transmissions of multiple LEO satellites. More details about the feasibility of such cooperative modes can be referred to \cite{shang2023coverage,abdelsadek2023broadband,abdelsadek2022distributed}.}
in terms of coverage probability and achievable data rate.
With more than one satellite providing desired signals, the introduced system may significantly improve the coverage probability and achievable data rate of UTs on the earth under the scattering environment.

\section{System Model}
A CoMS joint transmission system comprising of $M$ satellite APs (SAPs) in the space and $K$ UTs on the ground is investigated.
We assume that SAPs are coordinated by a central BS which is either on the ground or on an existing SAP with enough power and computing capabilities \cite{shang2023coverage}.

\subsection{Network Model}
The signal propagation is affected by path-loss attenuation and small-scale fading.
Let the channel coefficient between $m$-th SAP and $k$-th UT to be $g_{mk}={\beta_{mk}}^{1/2}h_{mk}$. Using a log-distance model, the path-loss attenuation can be modeled as $\beta_{mk}={\beta_{0} d_{mk}}^{-\alpha}$ where $\beta_{0}$ is the path-loss at a reference distance, and $d_{mk}$ is the distance in between.
We consider the Rayleigh small-scale fading channel\footnote{This can be reasonable when it refers to a drastic and complex fading environment on the earth, such as in canyon, wooded mountains, and terrestrial-communications-unavailable urban environments during emergencies, where the received signals can be affected by severe multi-path distortion. More sophisticated fading models will be investigated in the future work.} for ease of analysis \cite{okati2020downlink}, i.e., $h_{mk} \sim \mathcal{CN}(0,1)$.
Similar to \cite{okati2020downlink,park2022tractable}, a typical UT, i.e. the $k$-th UT located at the top of the earth, is selected as a representative. 
All SAPs within the typical UT's connected range, as defined in the following section, simultaneously serve the typical UT, while those outside this range will bring interference to its received signal. 
We assume that UTs associated with the same SAP are assigned orthogonal sub-channels to avoid mutual interference in proximity.

Let $q_m$ be the normalized data symbol from the $m$-th SAP such that $\mathbb{E}\{ |q_m|^2\} = 1$, and denote $\Phi_{k}$, $\Phi_{k}^{S}$, and $\Phi_{k}^{I}$ as the sets of all SAPs above UT's horizon line, serving SAPs, and interfering SAPs, respectively. 
As shown in Fig. 1, serving SAPs are within the connected range, and the remaining SAPs above the typical UT's horizon are interfering SAPs.
The signals received by the $k$-th UT is given by
\begin{equation}
    \begin{aligned}
        r_k &= \sqrt{P_d G_t} \sum_{m\in\Phi_{k}} {\beta_{mk}}^{1/2} h_{mk} q_{m} + w_{k} \\
            &= \begin{aligned}[t]
                    &\underbrace{ \sqrt{P_d G_t} \sum_{m\in\Phi_{k}^{S}} {\sqrt{\beta_{0}} d_{mk}}^{-\alpha/2} h_{mk} q_m }_{\mathrm{DS}_k}  \\
                    &+\underbrace{ \sqrt{P_d G_t} \sum_{m\in\Phi_{k}^{I}} {\sqrt{\beta_{0}} d_{mk}}^{-\alpha/2} h_{mk} q_{m} }_{\mathrm{IS}_k} + w_{k},
                \end{aligned}
        \label{Formula:received_signal_at_UT}
    \end{aligned}
\end{equation}
where $P_d$ is the normalized satellite transmit power, 
$G_t$ is the satellite transmit antenna gain, 
and $w_{k} \sim \mathcal{CN}(0,1)$ is the additive white Gaussian noise (AWGN). Moreover, $\mathrm{DS}_k$ and $\mathrm{IS}_k$ represent the desired signal and the interfering signal received at the $k$-th UT.

\subsection{Spatial Distribution}
We model the randomly deployed SAPs on the surface of a sphere in $\mathbb{R}^3$ centering at the origin $\mathbf{0} \in \mathbb{R}^3$, with radius $R_S$. All SAPs are distributed according to a homogeneous spherical Poisson point process (SPPP). The probability of having $n$ nodes within the sphere $\mathbb{S}_{R_S}^2$ is given by
\begin{equation}
    \mathbb{P} \left\{N (\mathbb{S}_{R_S}^2)=n \right\} = \frac{(4\pi {R_S}^2 \lambda_S)^n}{n!} \mathrm{exp}\left({-4\pi {R_S}^2 \lambda_S}\right),
    \label{Formula: SPPP}
\end{equation}
\captionsetup{font={scriptsize}}
\begin{figure}[ht]
\begin{center}
\setlength{\abovecaptionskip}{+0.2cm}
\setlength{\belowcaptionskip}{-0.6cm}
\centering
  \includegraphics[width=3.2in, height=1.6in]{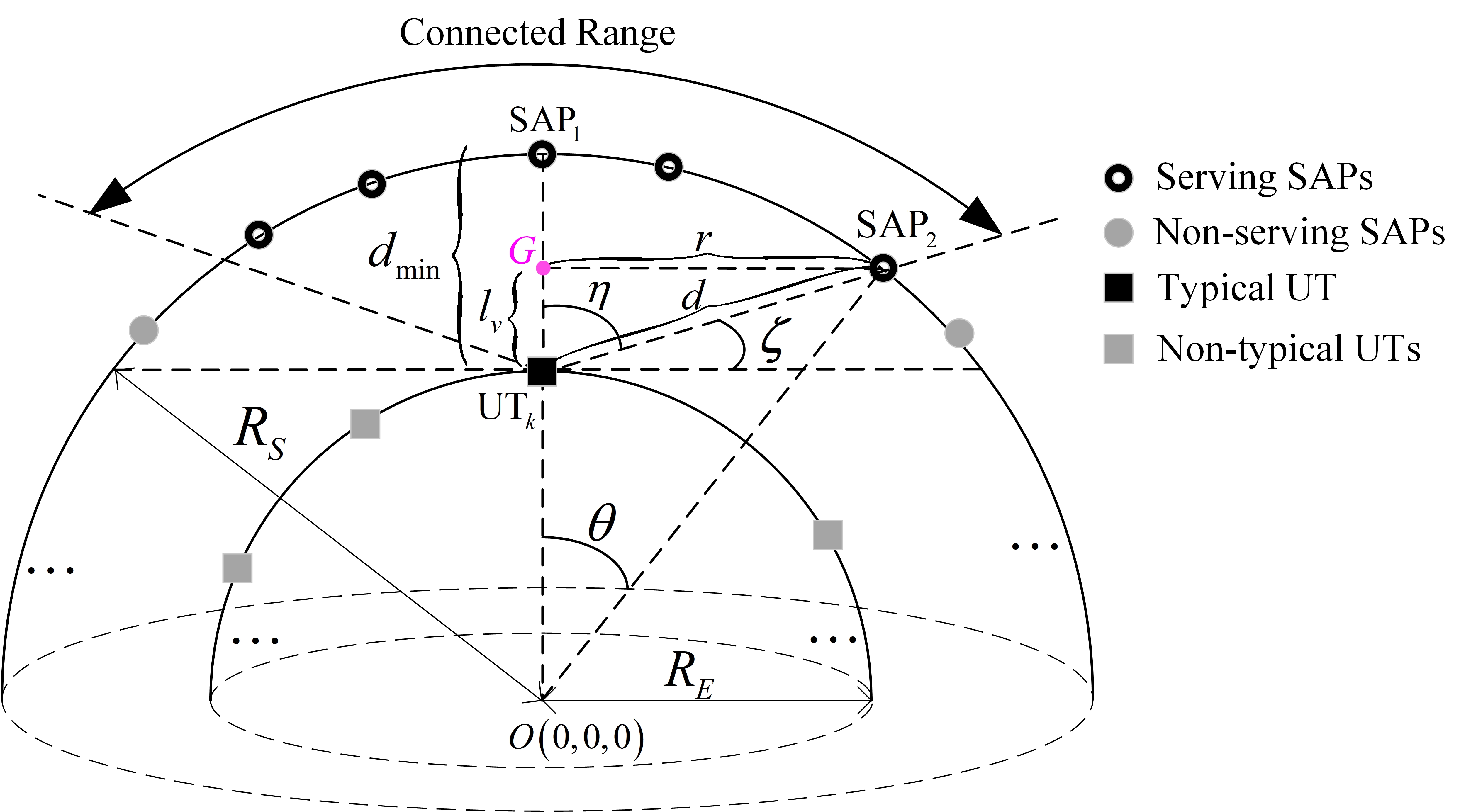}
\renewcommand\figurename{Fig}
\caption{\scriptsize A sketch of CoMS joint transmission system.}
\label{fig:Appendix_sphere}
\end{center}
\end{figure}
where $N (\mathbb{S}_{R_S}^2)$ denotes the number of SAPs in the sphere $\mathbb{S}_{R_S}^2$ and $\lambda_S$ is the density of SAPs. Similarly, define the surface of the earth to be $\mathbb{S}_{R_E}^2$ with radius $R_E(<R_S)$. 
All UTs are also randomly distributed on the surface with density $\lambda_U$ following the homogeneous SPPP. 
Moreover, the distribution of the SAPs and the UTs are independent on each other due to the movement of LEO satellites and UTs.

\section{Performance Analysis}
\subsection{Statistical Properties}
A sketch of modeling and related parameters for analyzing the typical UT have been depicted in Fig. \ref{fig:Appendix_sphere}. The typical UT is simultaneously served by SAPs in its connected range. The angle $\zeta$ represents the elevation angle from the typical UT's horizon line, while $\eta$ is its complementary angle in the same plane. The connected range is the area in cone angle of $2\eta$.

\begin{lemma}
The cumulative distribution function (CDF) of the distance $D$ from any one of the serving SAPs to the typical UT with the angle $\eta$ is given in (\ref{Formula:CDF_d}) at the top of next page,
\begin{figure*}[t]
\setlength{\abovecaptionskip}{-1.5cm}
\setlength{\belowcaptionskip}{-0.5cm}
\normalsize
\begin{equation}
\begin{split}
  F_{D}(d) = \begin{cases} 0, & 0<d<(R_S-R_E) \\ \frac{d^2-(R_S-R_E)^2}{2R_E(R_S-R_E \sin^2\eta - \sqrt{ {R_S}^2 -{R_E}^2\sin^2\eta } \cos\eta)}, & (R_S-R_E)\leq d \leq \sqrt{ {R_S}^2 -{R_E}^2\sin^2\eta }-R_E\cos\eta \\1, & d>\sqrt{ {R_S}^2 -{R_E}^2\sin^2\eta }-R_E\cos\eta \end{cases}
    \label{Formula:CDF_d}
\end{split}
\end{equation}
\hrulefill
\end{figure*}
where $\eta \in [0,\frac{\pi}{2}]$, and the corresponding probability density function (PDF) is given by
\begin{equation}
    f_D(d) = \frac{d}{R_E(R_S-R_E \sin^2\eta - \sqrt{ {R_S}^2 -{R_E}^2\sin^2\eta } \cos\eta)},
    \label{Formula:PDF_d}
\end{equation}
for $(R_S-R_E)\leq d \leq \sqrt{ {R_S}^2 -{R_E}^2\sin^2\eta }-R_E\cos\eta$ while $f_D(d) = 0$ otherwise.   
\end{lemma}

\textit{Proof}: The distance from any one of the serving SAPs to the typical UT in terms of the angle $\eta$ needs first to be given. In Fig. \ref{fig:Appendix_sphere}, consider one triangle formed by SAP$_2$, point $G$ and UT$_k$, and the other by SAP$_2$, point $G$ and original point $O$. Among known parameters $R_S$, $R_E$, $\eta$ and $\theta$, by using geometrical relationships, i.e. $r=R_S\sin\theta=d\sin\eta$ and $R_S\cos\theta=R_E+d\cos\eta$, the distance can be represented as
\begin{equation}
    d = \sqrt{{R_S}^2 - {R_E}^2 \sin^2\eta} - R_E\cos\eta.
    \label{Formula: d}
\end{equation}
Moreover, the CDF of the distance $D$ from any one of SAP on the outer sphere to the typical UT is provided in \cite{okati2020downlink} as
\begin{equation}
    F_D(d) = \begin{cases} 0, & 0<d<R_E \\ \frac{d^2-\left(R_S-R_E\right)^2}{4R_E R_S}, & R_E<d<(R_E+R_S) \\ 1, & d>(R_E+R_S) \end{cases}
    \label{Formula: CDF_in_Downlink}
\end{equation}
As the distance distribution is considered only for satellites above the horizon line of the typical UT, the middle-case CDF expression in (\ref{Formula: CDF_in_Downlink}) is multiplied with a scaling factor $\varpi = \frac{2R_S}{R_S-R_E\sin^2\eta -\cos\eta\sqrt{{R_S}^2-{R_E}^2\sin^2\eta}}$, the CDF in (\ref{Formula:CDF_d}) can be obtained and corresponding PDF is achieved via derivation over $d$, which completes the proof. \hfill$\square$

\begin{lemma}
The lowest altitude for the serving SAPs of the typical UT is given by
\begin{equation}
    l_v(\eta) = \begin{cases} \cos^2\eta \left( \sqrt{{R_E}^2 + \frac{{R_S}^2-{R_E}^2}{\cos^2\eta}}-R_E \right), & 0 \leq \eta < \frac{\pi}{2} \\ 0, & \eta=\frac{\pi}{2} \end{cases}
    \label{Formula:height}
\end{equation}
Then, the maximum distance from a serving SAP to the typical UT is $\frac{l_v}{\cos\eta}$ for $0 \leq \eta < \frac{\pi}{2}$, and $\sqrt{{R_S}^2-{R_E}^2}$ for $\eta=\frac{\pi}{2}$.
\end{lemma}

\textit{Proof}: For previous two triangles, by taking trigonometric function $l_v=r\tan\eta$ and Pythagorean theorem ${R_S}^2=r^2+(l_v+R_E)^2$, the lowest altitude for serving SAPs is given as
\begin{equation}
    l_v(\eta) = \cos^2\eta \left( \sqrt{{R_E}^2 + \frac{{R_S}^2-{R_E}^2}{\cos^2\eta}}-R_E \right),
\end{equation}
for $0 \leq \eta < \frac{\pi}{2}$, while $l_v(\eta)=0$ is obvious for $\eta = \frac{\pi}{2}$, which completes the proof. \hfill$\square$

\begin{lemma}
The average number of ground UTs within the serving area of an SAP is given by
\begin{equation}
    N_U = 2\pi R_E\lambda_U \begin{aligned}[t]
    &\left(R_E-\frac{{R_E}^2}{R_S} \sin^2\eta \right.\\
    &\left. - \frac{R_E\sqrt{{R_S}^2-{R_E}^2\sin^2\eta}\cos\eta}{R_S} \right).
    \end{aligned}
    \label{Formula: number_of_UTs}
\end{equation}
\end{lemma}

\begin{figure}[ht]
\begin{center}
\setlength{\abovecaptionskip}{+0.2cm}
\setlength{\belowcaptionskip}{-0.2cm}
\centering
  \includegraphics[width=2.6in, height=1.6in]{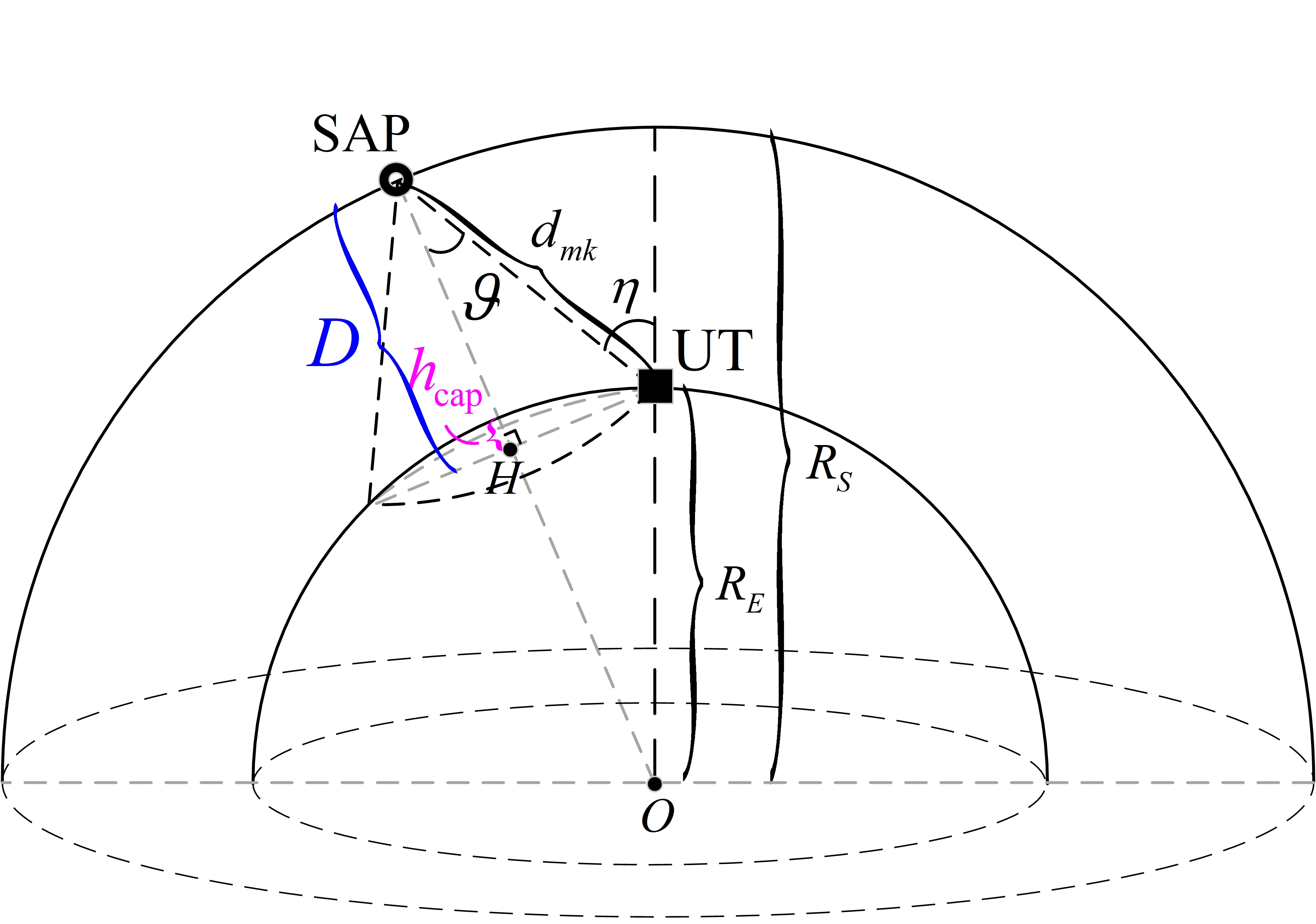}
\renewcommand\figurename{Fig}
\caption{\scriptsize Statistical Properties for coverage area of a satellite.}
\label{fig:Coverage_area}
\end{center}
\end{figure}

\textit{Proof}: As shown in Fig. \ref{fig:Coverage_area}, the typical UT is at the top of the inner semicircle which represents the surface of the earth, while a certain SAP is on the outer semicircle which represents LEO. The maximum distance from a serving SAP of the typical UT to its coverage area is given in (\ref{Formula: d}) of Lemma 1, as has been marked as $d_{mk}$ in the figure.

Consider the triangle formed by the SAP, the typical UT and origin $O$. From the cosine theorem, we have the relationship $\cos\vartheta = \frac{{R_S}^2-{R_E}^2+{d}^2}{2R_S d}$. In another triangle formed by the SAP, the typical UT and point $H$, by inserting $d$ and $\cos\vartheta$, we have distance $D= d\cos\vartheta$ in the figure given by
\begin{equation}
\begin{aligned}
    D &= R_S - \frac{{R_E}^2}{R_S}\sin^2\eta - \frac{R_E\sqrt{{R_S}^2 - {R_E}^2\sin^2\eta}\cos\eta}{R_S}.
\end{aligned}
\end{equation}
The height of the spherical cap can be written as $h_{\mathrm{cap}} = D + R_E - R_S$. Using surface area formula for the spherical cap $S_{\mathrm{service}} = 2\pi R_E h_{\mathrm{cap}}$, the average number of UTs in coverage of SAP is written as in (\ref{Formula: number_of_UTs}), which completes the proof. \hfill$\square$
\captionsetup{font={scriptsize}}

\subsection{Coverage and Rate Analysis}
\hspace{-3.5mm}\textbf{Definition 1}: If the downlink signal-to-interference-plus-noise-ratio (SINR) from its serving SAPs is higher than the threshold value $\gamma_{th}$, this typical UT is in the coverage. With the received signal in (\ref{Formula:received_signal_at_UT}), the SINR of the typical UT is given by
\begin{equation}
    \mathrm{SINR}_k = \frac{P_d G_t \beta_{0} \left|\sum_{m\in\Phi_{k}^{S}} {d_{mk}}^{-\alpha/2} h_{mk} \right|^2}{P_d G_t \sum_{m\in\Phi_{k}^{I}} \beta_{0} {d_{mk}}^{-\alpha} \left|h_{mk}\right|^2 + {\sigma_{n_0}}^2},
\end{equation}
where ${\sigma_{n_0}}^2$ is the variance of AWGN. The coverage probability of the typical UT is written as
\begin{equation}
    \begin{aligned}
        & \mathbb{P}_k^{\mathrm{cov}}(\gamma_{th};\lambda_S,\lambda_U,R_S,P_d,G_t) \\
        &= \mathbb{P} \left\{ \mathrm{SINR}_k>\gamma_{th} \right\} \\
        &= \mathbb{P} \left\{ \frac{P_d G_t \beta_{0} \left|\sum_{m\in\Phi_{k}^{S}} {d_{mk}}^{-\alpha/2} h_{mk} \right|^2}{P_d G_t \sum_{m\in\Phi_{k}^{I}} \beta_{0} {d_{mk}}^{-\alpha} \left|h_{mk}\right|^2 + {\sigma_{n_0}}^2} > \gamma_{th} \right\} \\
         &= \mathbb{P} \begin{aligned}[t]
                 & \left\{ \left|\sum_{m\in\Phi_{k}^{S}} {d_{mk}}^{-\alpha/2} h_{mk} \right|^2 \right.\\ 
                &\left. > \gamma_{th} \sum_{m\in\Phi_{k}^{I}} {d_{mk}}^{-\alpha} \left|h_{mk}\right|^2 + \frac{\gamma_{th} {\sigma_{n_0}}^2}{P_d G_t \beta_{0}} \right\}.
            \end{aligned}       
    \end{aligned}
\end{equation}


\begin{theorem}
The coverage probability of the system is given in (\ref{Theorem1}) at the top of next page.
\begin{figure*}[t]
\setlength{\abovecaptionskip}{-1.5cm}
\setlength{\belowcaptionskip}{-0.6cm}
\normalsize
\begin{equation}
\begin{split}
  \begin{array}{l}
  \mathbb{P}_k^{{\rm{cov}}}({\gamma _{th}};{\lambda _S},{\lambda _U},{R_S},{P_d},{G_t}) \approx {\rm{exp}}\left( { - \pi {\lambda _S}\frac{{{R_S}}}{{{R_E}}}{s^{\frac{2}{\alpha}}}\int_{s^{-\frac{2}{\alpha}}{r_{I,{\rm{min}}}}^2}^{s^{-\frac{2}{\alpha}}{r_{I,{\rm{max}}}}^2} 1  - \frac{1}{{1 + {u^{ - \alpha /2}}}}du - \frac{{s{\sigma _{{n_0}}}^2}}{{{P_d}{G_t}{\beta _0}}}} \right)\\
  {\rm{where }}\quad
  {\rm{ }} s=\lambda_R \gamma_{th}, \\
  {r_{I,{\rm{min}}}} = \sqrt {{R_S}^2 - {R_E}^2{{\sin }^2}\eta }  - {R_E}\cos \eta,
  \quad {r_{I,{\rm{max}}}} = \sqrt {{R_S}^2 - {R_E}^2},\\
  {r_{S,{\rm{min}}}} = {R_S} - {R_E},
  \quad {r_{S,{\rm{max}}}} = \sqrt {{R_S}^2 - {R_E}^2{{\sin }^2}\eta }  - {R_E}\cos \eta.
\end{array}
    \label{Theorem1}
\end{split}
\end{equation}
\hrulefill
\vspace{-2mm}
\end{figure*}
\end{theorem}

\textit{Proof}: Consider the typical spherical cap $\mathcal{A}_r$ in \cite{park2022tractable}, i.e.  
\begin{equation}
    \left| \mathcal{A}_r \right|= 2\pi\left[ R_S-R_E-\frac{\left( {R_S}^2 - {R_E}^2 \right)-r^2}{2R_E} \right] R_S,
\end{equation}
where $r$ is the distance from an SAP to the typical UT. Denote $\frac{1}{\lambda_R}=\sum_{m\in\Phi_{k}^{S}} {d_{mk}}^{-\alpha}$ as the sum of channels between the typical UT and its serving SAPs. Taking the derivation $\frac{\partial \left| \mathcal{A}_r \right|}{\partial r} = 2\frac{R_S}{R_E}\pi r$, we further represent the sum of channels as
\begin{equation}
    \begin{aligned}
        \mathbb{E}\left\{ \frac{1}{\lambda_R} \right\} 
        &= \mathbb{E}\left\{ \sum_{m\in\Phi_{k}^{S}} {d_{mk}}^{-\alpha} \right\} 
        = \lambda_S \frac{\partial \left| \mathcal{A}_r \right|}{\partial r} \int_{r_{S,\mathrm{min}}}^{r_{S,\mathrm{max}}} r^{-\alpha} dr \\
        &= 2\frac{R_S}{R_E} \pi\lambda_S \int_{r_{S,\mathrm{min}}}^{r_{S,\mathrm{max}}} r^{1-\alpha} dr,
    \end{aligned}
    \label{Formula: 1/lambda_R}
\end{equation}
where $r_{S,\mathrm{min}} = R_S - R_E$, $r_{S,\mathrm{max}} = \sqrt{R_S^2 -R_E^2\sin^2\eta} - R_E\cos\eta$. Then, by denoting $I_k = \sum_{m\in\Phi_{k}^{I}} {d_{mk}}^{-\alpha} \left|h_{mk}\right|^2$, the coverage probability can be written as
\begin{equation}
    \begin{aligned}
        &\mathbb{P}_k^{\mathrm{cov}}(\gamma_{th};\lambda_S,\lambda_U,R_S,P_d,G_t) \\
        &= \mathbb{P}_k \left\{ \left|\sum_{m\in\Phi_{k}^{S}} {d_{mk}}^{-\frac{\alpha}{2}} h_{mk} \right|^2 > \gamma_{th} I_k + \frac{\gamma_{th} {\sigma_{n_0}}^2}{P_d G_t \beta_{0}} \right\} \\
        &\stackrel{(a)}{=} \mathbb{E}_{\Phi_{k}^{I}} \left\{ \mathrm{exp} \left[ -\frac{1}{\sum_{m\in\Phi_{k}^{S}} {d_{mk}}^{-\alpha}} \left( \gamma_{th} I_k + \frac{\gamma_{th} {\sigma_{n_0}}^2}{P_d G_t \beta_{0}} \right) \right]  \right\} \\
        &\stackrel{(b)}{\approx} \mathbb{E}_{\Phi_{k}^{I}} \left\{ \mathrm{exp} \left[ -\lambda_R \left( \gamma_{th} I_k + \frac{\gamma_{th} {\sigma_{n_0}}^2}{P_d G_t \beta_{0}} \right) \right]  \right\} \\
        &\stackrel{(c)}{=} \mathbb{E}_{\Phi_{k}^{I}} \left\{ \mathcal{L}_{I_k}\left( s \right) \mathrm{exp} \left( -\frac{s {\sigma_{n_0}}^2}{P_d G_t \beta_{0}} \right) \right\},
    \end{aligned}
    \label{Formula: Coverage probability proof}
\end{equation}
where $(a)$ follows from $\left| \sum_{m\in\Phi_{k}^{S}} {d_{mk}}^{-\alpha/2} h_{mk} \right|^2 \sim \mathrm{exp}(\lambda_R)$; in $(b)$ the approximation holds due to (\ref{Formula: 1/lambda_R}); $(c)$ holds true because of the Laplace transform (LT) of the interference, i.e., $\mathcal{L}_I(s)=\mathbb{E}\left\{e^{-sI}\right\}$, and let $s=\lambda_R\gamma_{th}$. For a given $\gamma_{th}$, denote $s=\lambda_R \gamma_{th}$, the LT of the interference is given by
\begin{equation}
    \begin{aligned}
        &\mathcal{L}_{I_k}(s)= \mathbb{E}_{\Phi_{k}^{S},h} \left\{ \mathrm{exp}(-s I_k) \right\} \\
        &= \mathbb{E}_{\Phi_{k}^{S}} \left\{ \prod_{m\in\Phi_{k}^{I}} \mathbb{E}_{h} \left\{ \mathrm{exp} \left( -s \left|h_{mk}\right|^2 {d_{mk}}^{-\alpha} \right) \right\} \right\} \\
        &= \mathbb{E}_{\Phi_{k}^{S},\left|h \right|^2} \left\{ \prod_{m\in\Phi_{k}^{I}} \frac{1}{1+s {d_{mk}}^{-\alpha}} \right\} \\
        &= \mathrm{exp} \left( -\lambda_S \int_{d\in\mathcal{A}_{k}^{I}}1-\frac{1}{1+s {d_{mk}}^{-\alpha}}dr \right) \\
        &= \mathrm{exp} \left( -\pi\lambda_S \frac{R_S}{R_E} s^{\frac{2}{\alpha}} \int_{s^{-\frac{2}{\alpha}}{r_{I,\mathrm{min}}}^2}^{s^{-\frac{2}{\alpha}}{r_{I,\mathrm{max}}}^2} 1-\frac{1}{1+u^{-\frac{\alpha}{2}}}du \right), \\
    \end{aligned}
    \label{Formula: Laplace transform}
\end{equation}
where $r_{I,\mathrm{min}} = \sqrt{{R_S}^2 - {R_E}^2\sin^2\eta} - R_E\cos\eta$, $r_{I,\mathrm{max}} = \sqrt{{R_S}^2 - {R_E}^2}$. Invoke (\ref{Formula: Laplace transform}) into (\ref{Formula: Coverage probability proof}), we finish the proof. \hfill$\square$

\begin{theorem}
The achievable data rate of the system is
\begin{equation}
    R_k = \frac{B}{N_{U}} \int_{0}^{\infty} \begin{aligned}[t]
    &{\rm{exp}}\left(  - \pi {\lambda _S}\frac{{{R_S}}}{{{R_E}}}{s^{\frac{2}{\alpha }}}\int_{s^{-\frac{2}{\alpha }}{r_{I,{\rm{min}}}}^2}^{{s^{-\frac{2}{\alpha }}r_{I,{\rm{max}}}}^2} \right.\\ 
            &\left. 1  - \frac{1}{{1 + {u^{ - \alpha /2}}}}du - \frac{{s{\sigma _{{n_0}}}^2}}{{{P_d}{G_t}{\beta _0}}} \right) dt,
    \end{aligned}
\end{equation}
\textit{where $B$ is the bandwidth, $s = \lambda_R(2^t-1)$, and $N_U$ is the average number of UTs served by each SAP given in (\ref{Formula: number_of_UTs}).}
\end{theorem}

\textit{Proof}: Since $\mathbb{E}\{X\}=\int_{0}^{\infty}\mathbb{P}\{X>x\}dx$ for $X>0$,
\begin{equation}
    \begin{aligned}
        \mathbb{E}\left\{ \mathrm{log}_2 (1+\mathrm{SINR}_k) \right\}
        &= \int_{0}^{\infty} \mathbb{P} \left\{ \mathrm{SINR}_k > 2^t-1 \right\} dt.
    \end{aligned}
\end{equation}
Take $\gamma_{th} = 2^t-1$ and insert (\ref{Theorem1}), we complete this proof. \hfill$\square$

\section{Simulations and Numerical Results}
In simulations, a number of UTs are scattered on the surface of the earth with a radius of $6,371.393$ km while SAPs are distributed on the orbit of $500$ km above the earth, both according to SPPP. The densities of UTs and SAPs are $10^{-6}$ per km\textsuperscript{2} and $5\times10^{-6}$ per km\textsuperscript{2}, respectively. The coverage probability comparisons are obtained by using Monte Carlo simulations for $10,000$ times. The path-loss factor $\alpha=2$.

\begin{figure}[t]
\begin{center}
\setlength{\abovecaptionskip}{+0.2cm}
\setlength{\belowcaptionskip}{-0.7cm}
\centering
  \includegraphics[width=2.8in, height=1.73in]{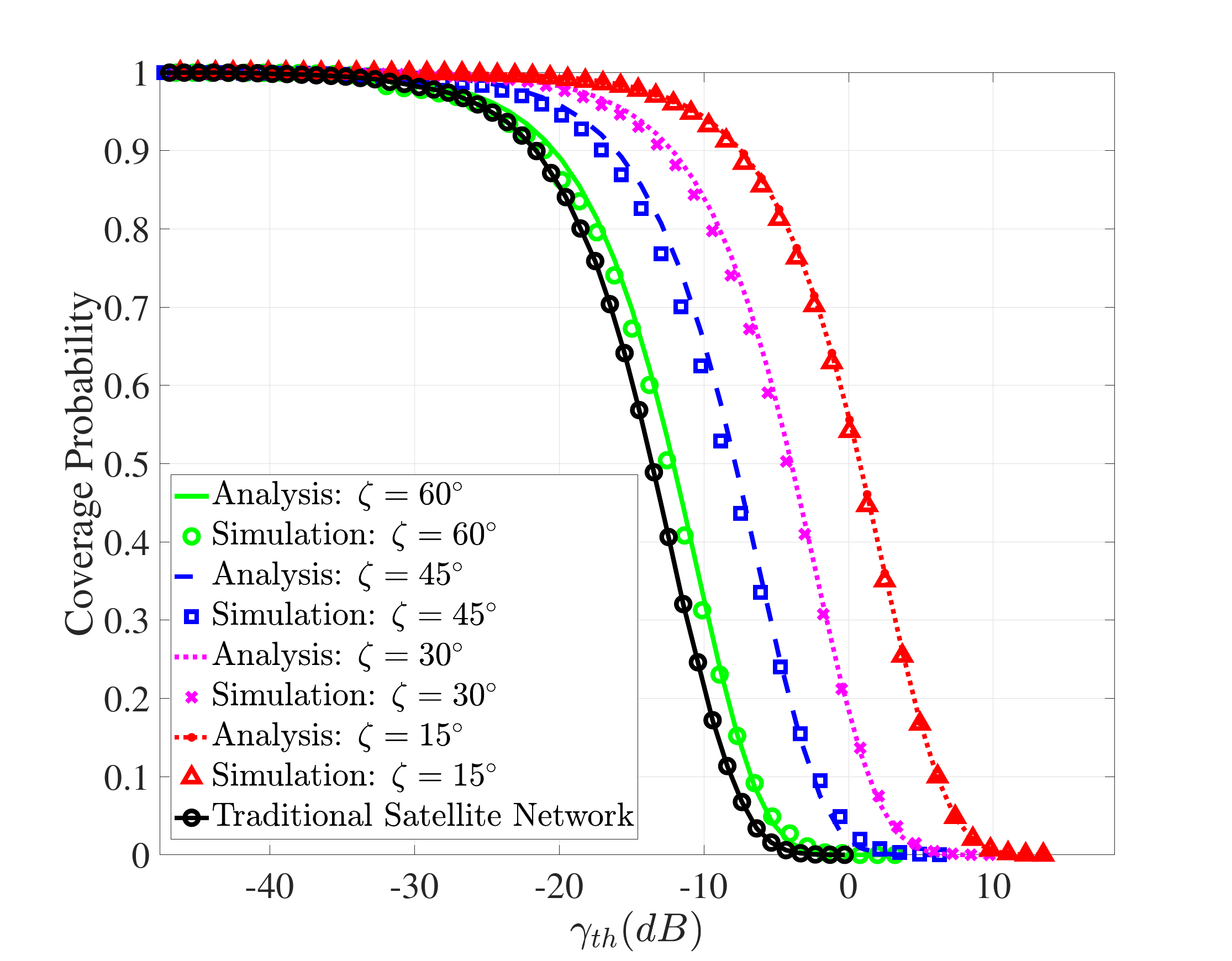}
\renewcommand\figurename{Fig}
\caption{\scriptsize Coverage probability with different elevation angles and comparison with traditional satellite network.}
\label{fig:Coverage_Probability}
\end{center}
\end{figure}

Fig. \ref{fig:Coverage_Probability} demonstrates the coverage probability for various threshold values from $-50$ dB to $20$ dB. 
Simulations verify our derived analytical results.
With a smaller elevation angle, UT can have a higher coverage probability under the same threshold value, which is in line with our intuition.
Moreover, these results are compared with those in the traditional satellite network where a typical UT is served by the nearest satellite, and better performances can be achieved by using the CoMS joint transmission system. 
This is especially meaningful for UTs in emergent circumstances when terrestrial networks are temporarily out of use. 

\captionsetup{font={scriptsize}}

\begin{figure}[t]
\begin{center}
\setlength{\abovecaptionskip}{+0.2cm}
\setlength{\belowcaptionskip}{-0.6cm}
\centering
  \includegraphics[width=2.8in, height=1.73in]{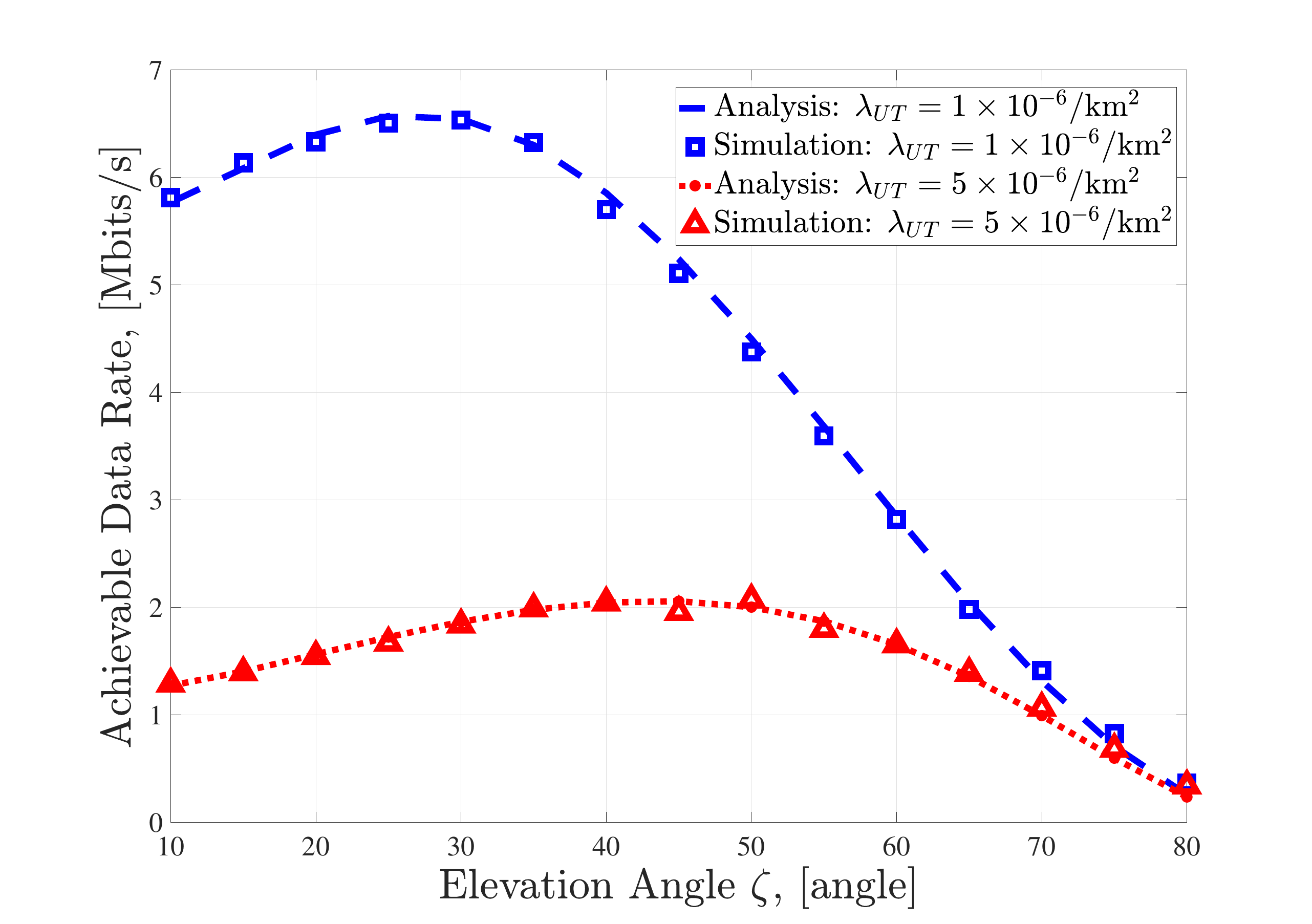}
\renewcommand\figurename{Fig}
\caption{\scriptsize Achievable data rate against elevation angle under different densities of UTs. The altitude and density of SAP are $h=500$ km, $\lambda_{SAP}=5\times 10^{-6}/ \mathrm{km^2}$, respectively.}
\label{fig:3_1_data_rate_density}
\end{center}
\end{figure}

Fig. \ref{fig:3_1_data_rate_density} compares the achievable downlink data rate under different densities of UTs. In each case there exists an optimal elevation angle for maximal achievable data rate. Along with the increase of UT density, comes the increase of the optimal elevation angle. The maximal achievable data rate also decreases when the UT density increases. 
As the elevation angle increases, a satellite' bandwidth is allocated to fewer UTs due to per-UT's lower visibility, which incurs more interference signals, fewer desired signals, and thus lower SINR. 
The bandwidth allocation influences the achievable data rate; however, the increase in the UT density results in the per-satellite bandwidth being allocated to more UTs, so that the positive contribution of the bandwidth to per-UT achievable data rate is generally reduced.
The trade-off between allocated bandwidth and SINR determines the optimal elevation angles.

\captionsetup{font={scriptsize}}

\begin{figure}[t]
\begin{center}
\setlength{\abovecaptionskip}{+0.2cm}
\setlength{\belowcaptionskip}{-0.7cm}
\centering
  \includegraphics[width=2.8in, height=1.73in]{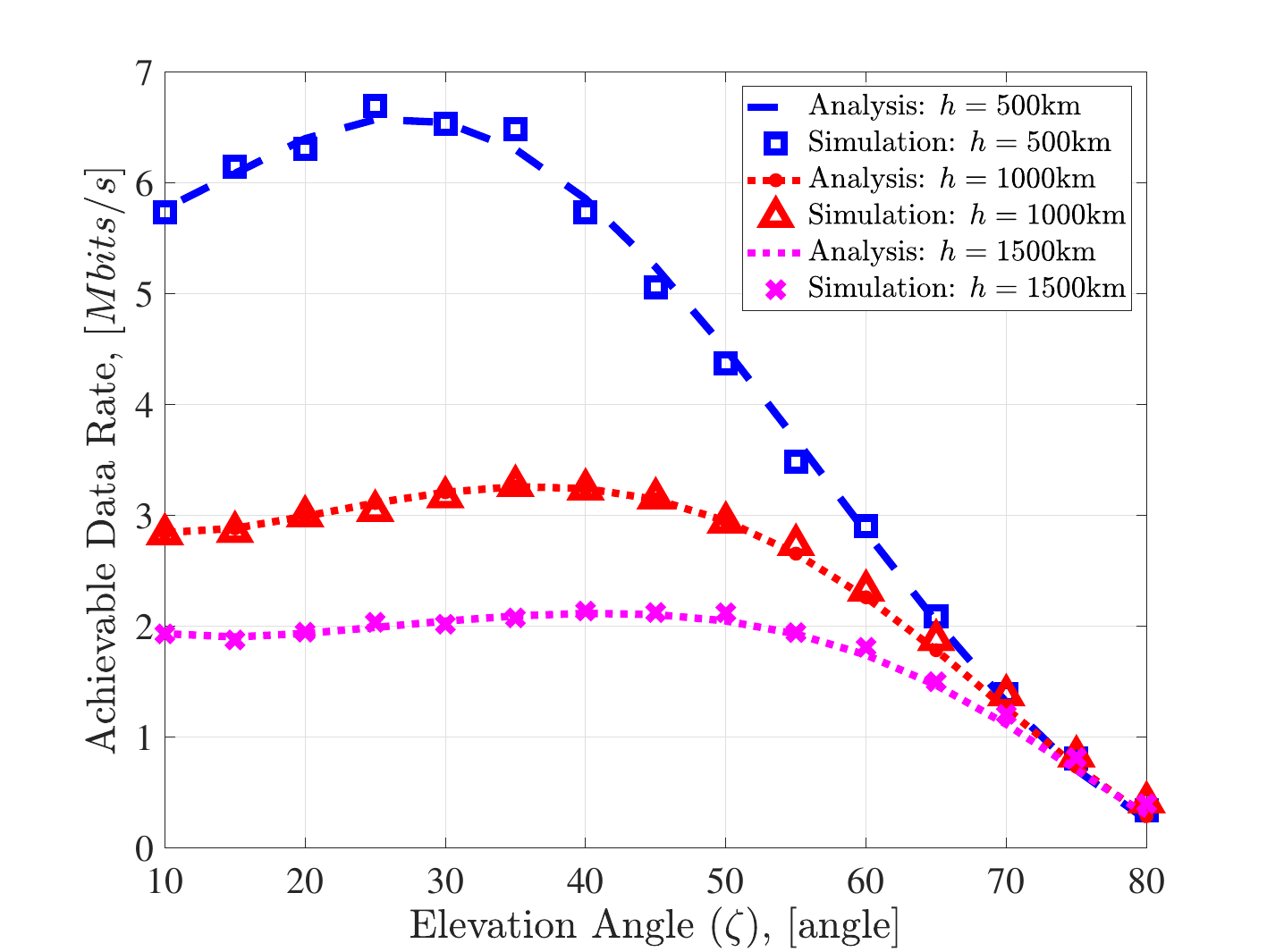}
\renewcommand\figurename{Fig}
\caption{\scriptsize Achievable data rate against elevation angle under various SAPs' altitudes. Densities of UT and SAP are $1\times 10^{-6}$ / km\textsuperscript{2} and $5\times 10^{-6}$ / km\textsuperscript{2}, respectively.}
\label{fig:4_1_data_rate_altitude}
\end{center}
\end{figure}

The achievable downlink data rate for different altitudes of SAPs is shown Fig. \ref{fig:4_1_data_rate_altitude}. The altitude of SAPs has a large influence on both achievable data rate and optimal elevation angle. For instance, when SAP altitude is $500$ km, the maximal achievable data rate is $6.6$ Mbits/s under $\zeta = 25^{\circ}$. The maximal achievable data rate drops to $3.3$ Mbits/s under $\zeta = 35^{\circ}$ when SAP altitude is $1,000$ km. 
The optimal elevation angle becomes less evident as the SAP altitude increases.

\captionsetup{font={scriptsize}}

\begin{figure}[t]
\begin{center}
\setlength{\abovecaptionskip}{+0.2cm}
\setlength{\belowcaptionskip}{-0.6cm}
\centering
  \includegraphics[width=2.8in, height=1.73in]{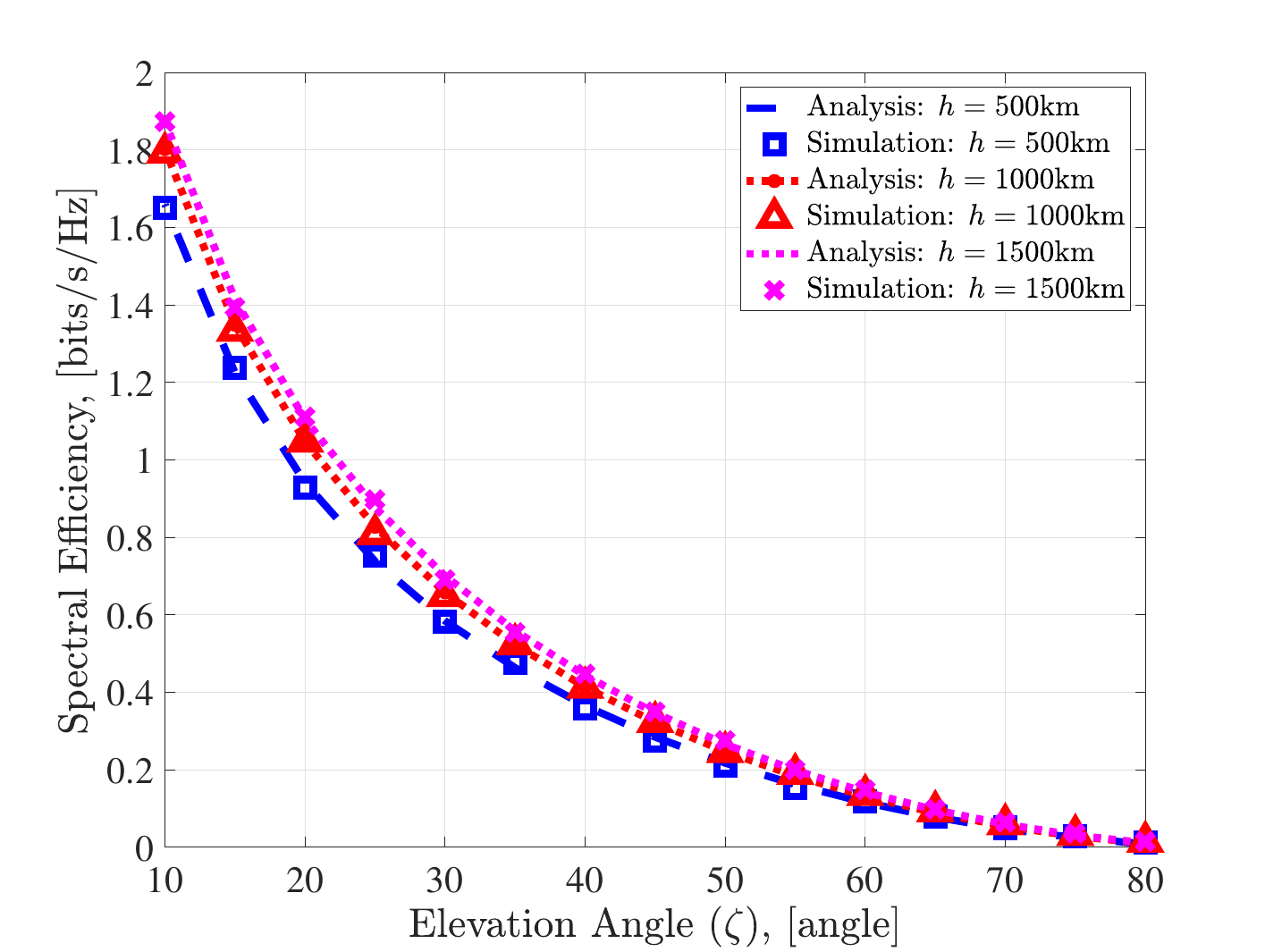}
\renewcommand\figurename{Fig}
\caption{\scriptsize Spectral efficiency against elevation angle under different altitudes of SAPs, where $\lambda_{UT}=1\times 10^{-6}$ / km\textsuperscript{2} and $\lambda_{SAP}=5\times 10^{-6}$ / km\textsuperscript{2}.}
\label{fig:4_2_SE_altitudey}
\end{center}
\end{figure}

In Fig. \ref{fig:4_2_SE_altitudey}, the spectral efficiency (SE) of the typical UT is also compared under different altitudes of SAPs. It is observed that SE increases with a higher SAP altitude. 
This is because the decrease of interference power dominates the increase of desired signal power.
The data rate decreases with a higher altitude due to the reduction in bandwidth for UT.
\captionsetup{font={scriptsize}}

\section{Conclusion}
In this paper, the CoMS joint transmission system was investigated.
We analyzed the performance by considering the elevation angles, satellite altitudes and distribution densities of both SAPs and UTs. 
Simulations and numerical results show that the introduced system improves the coverage 2-15 dB compared to the traditional nearest-satellite-supported network.
Moreover, the higher altitude of SAPs brings the lower achievable data rate whilst the higher SE given the total number of satellites. 
In addition, an optimal elevation angle can be obtained to maximize the average achievable data rate.

\section{Acknowledgement}
This work was supported in part by the YongRiver Scientific and Technological Innovation Project No. 2023A-187-G.

\bibliographystyle{IEEEtran}
\bibliography{references.bib}

\end{document}